\documentclass[a4paper,fleqn]{mnras}

\usepackage{newtxtext,newtxmath}

\usepackage[T1]{fontenc}
\usepackage{ae,aecompl}

\usepackage{graphicx}	
\usepackage{amsmath}	
\usepackage{amssymb}	
\usepackage{subcaption}
\def\sigsfr{$\Sigma_{\rm SFR}$}
\def\sigmass{$\Sigma_{\star}$}
\def\dsigsfr{$\Delta \Sigma_{\rm SFR}$}

\title[SFRs and metallicities in mergers from MaNGA]{Spatially resolved star formation and metallicity profiles in post-merger galaxies from MaNGA}

\author[M. D. Thorp et al.]{
Mallory D. Thorp,$^{1}$\thanks{E-mail: mallorythorp@uvic.ca}
Sara L. Ellison,$^{1}$
Luc Simard,$^{2,1}$
Sebastian F. S\'anchez,$^{3}$
\newauthor
Braulio Antonio$^{1,3}$
\\
$^{1}$Department of Physics \& Astronomy, University of Victoria, Finnerty Road, Victoria, British Columbia, V8P 1A1, Canada\\
$^{2}$National Research Council of Canada, Herzberg Institute of Astrophysics, 5071 West Saanich Road, Victoria, British Columbia, Canada\\
$^{3}$Instituto de Astronom{\'i}a, Universidad Nacional Autonoma de Mexico, A.P. 70-264, C.P. 04510, Mexico, D.F., Mexico
}

\begin{document}
\label{firstpage}
\pagerange{\pageref{firstpage}--\pageref{lastpage}}
\maketitle

\begin{abstract} 
  Large galaxy surveys have demonstrated that galaxy-galaxy mergers can dramatically change
the morphologies, star formation rates (SFRs) and metallicities of their constituents. However,
most statistical studies have been limited to the measurement of global quantities, through large
fibres or integrated colours. In this work, we present the first statistically significant study
of spatially resolved star formation and metallicity profiles using integral field spectroscopy, using a sample of $\sim$20,000 spaxels in 36 visually selected post-merger galaxies from the Mapping Nearby Galaxies with Apache Point Observatory (MaNGA) survey. By measuring offsets from SFR and metallicity scaling relations on a spaxel-by-spaxel basis, we are able to quantify where in the galaxy these properties are most affected by the interaction. We find that the SFR enhancements are generally centrally peaked, by a factor of 2.5 on average, in agreement with predictions from simulations. However, there is considerable variation in the SFR behaviour in the galactic outskirts, where both enhancement and suppression are seen. The median SFR remains enhanced by 0.1 dex out to at least 1.9 Re. The metallicity is also affected out to these large radii, typically showing a suppression of $\sim -0.04$ dex.
\end{abstract}


\section{Introduction}

Simulations have long predicted that galaxy-galaxy interactions can have a dramatic effect on morphologies, star formation rates (SFRs), metallicities and nuclear accretion (e.g. Toomre \& Toomre 1972; Barnes \& Hernquist 1996; Mihos \& Hernquist 1996; Springel, Di Matteo \& Hernquist 2005; Di Matteo et al. 2008).  Observational studies have a similarly long history of confirming these predictions (e.g. Larson \& Tinsley 1978; Keel et al. 1985; Kennicutt et al. 1987).  Contemporary galaxy surveys, that include thousands of mergers, have put these empirical results onto a firm statistical footing, allowing the investigation of physical properties as a function of mass, mass ratio, merger stage and environment (e.g. Ellison et al. 2010; Scudder et al. 2012).

Traditionally, the main physical mechanism considered to be responsible for triggering many of the changes (e.g. in SFR, metallicity and nuclear accretion) during an interaction has been the growth of non-axisymmetric structures, such as a bar (e.g. Barnes \& Hernquist 1996; Mihos \& Hernquist 1996), which can funnel gas towards the centre of the galaxy.  Drainage of gas from the outskirts, and potentially lower gas surface densities due to tidal effects, can also lead to a suppression of star formation in the outer parts of the interacting galaxies (e.g. Moreno et al. 2015).  The same gas inflows are predicted to lead to a re-distribution of metals, reducing the central gas-phase metallicity and producing an overall flatter abundance gradient (Montuori et al. 2010; Rupke et al. 2010a; Perez et al. 2011; Torrey et al. 2012; Sillero et al. 2017; Bustamante et al. 2018).  However, additional physical processes may also be at work, such as shocks, compressive tidal forces and turbulence which can affect star formation well beyond the nuclear region (e.g. Powell et al. 2013; Renaud et al. 2014).

In order to empirically test the relative importance of the various physical mechanisms at work in a galaxy-galaxy interaction, it is necessary to spatially map the SFR and metallicity within a given galaxy.  By comparing SFRs both inside and beyond fibres in the Sloan Digital Sky Survey (SDSS), Ellison et al. (2013) concluded that although SFR enhancements are largest in galactic centres, they persist even at large radii.  However, detailed mapping requires either multi-object spectroscopy (MOS, e.g. Kewley et al. 2010; Rupke et al. 2010b; Rosa et al. 2014) or integral field units (IFUs, e.g. Garc{\'i}a-Mar{\'i}n et al. 2009; Rich et al. 2012; Cortijo-Ferrero et al. 2017).  Unfortunately, the sample sizes of most previous galaxy merger MOS and IFU studies have been small.  The advent of large IFU surveys is set to revolutionize our ability to spatially map galaxy properties for statistically significant samples of galaxy mergers.  For example, based on $\sim$100 galaxy mergers selected from the Calar Alto Legacy Integral Field Area (CALIFA, S{\'a}nchez et al. 2012) survey, Barrera-Ballesteros et al. (2015) studied the spatial changes in SFR and metallicity in two aperture bins (see also S{\'a}nchez et al. 2014).   However, full spatial profiling of SFRs and metallicities in mergers drawn from the new generation of IFU surveys has yet to be realized.

In the work presented here, we identify a sample of post-merger galaxies (i.e. merger remnants after the coalescence of two separate galaxies) from the Mapping Nearby Galaxies at Apache Point Observatory (MaNGA) IFU survey (Bundy et al. 2015) and measure the changes in their SFRs and metallicities on kpc-scales. This is the first statistically robust study of post-mergers to derive full radial profiles of these characteristics. We adopt a cosmology in which $H_0$ = 70 km s$^{-1}$ Mpc$^{-1}$, $\Omega_{M}$ = 0.3, $\Omega_{\Lambda}$ = 0.7.

\section{Methods}
\label{sec:methods}

\subsection{MaNGA Sample Selection}
\label{sec:samp}

The data release 14 (DR14) of the SDSS includes $\sim$2700 datacubes obtained for the  MaNGA survey whose IFUs are made of bundles of 2 arcsecond fibres. The MaNGA observing strategy ensures that galaxies are observed out to at least 1.5 $R_e$ and dithered observations, re-sampled to 0.5 arcsec spaxels, provide complete spatial coverage in the datacube (Law et al. 2015).

From the DR14 MaNGA sample, we have used the SDSS Sky Server $gri$ images ($r$-band half-light surface brightness limit of 23.0 mag arcsec$^{-2}$; Strauss et al. 2002) to visually identify 48 post-merger galaxies, defined as single galaxies with obvious tidal features such as tails and shells. 12 galaxies were omitted due to emission line cuts in the spaxel matching process (see Section 2.2), resulting in a 36 galaxy sample.

\begin{figure}
	\includegraphics[width=\columnwidth]{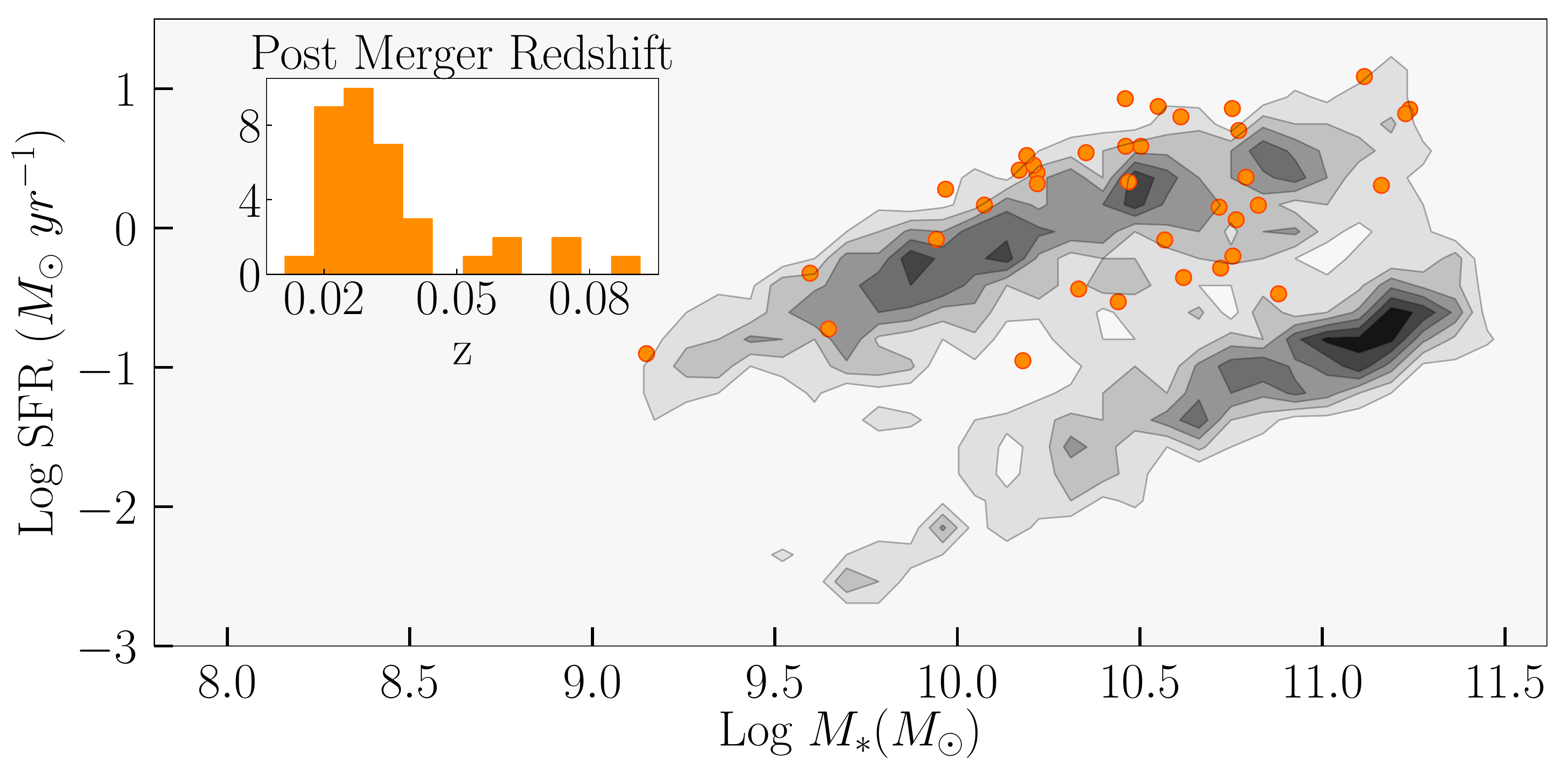}
	\centering
    \caption{SFR-mass distribution of the full DR14 MaNGA data set (grey contours), with our post-merger sample overlaid as orange points.  The inset shows the redshift distribution of the post-merger sample.}
    \label{fig:SFRvM}
\end{figure}

Our study uses publicly available data products obtained from MaNGA datacubes that have been processed using the {\sc pipe3d} spectral fitting pipeline\footnote{S{\'a}nchez et al. (2018), \hyperlink{http://www.sdss.org/dr14/manga/manga-data/manga-pipe3d-value-added-catalog/}{http://www.sdss.org/dr14/manga/manga-data/manga-pipe3d-value-added-catalog/}}, which is described in detail in S{\'a}nchez et al. (2016a, b).  Relevant to the current work, {\sc pipe3d} provides the star formation rate surface density (\sigsfr), stellar mass surface density (\sigmass), metallicity (O/H) and various emission line fluxes for each spaxel. The star formation rate is derived from the H$\alpha$ flux using the Kennicutt et al. (1998) relation and the gas phase metallicity is determined using the O3N2 calibration of Marino et al. (2013). The {\sc pipe3d} value added catalogue also provides integrated global properties including SFR and $M_{\star}$. These values were used, along with our redshift calculation, to compute global offsets in star formation rate from the star-forming main sequence ($\Delta$SFR). A distribution of SFR, $M_{\star}$, and redshift of our post-merger sample, with respect to the total MaNGA sample, is shown in Fig.\ref{fig:SFRvM}.

\subsection{Spaxel star formation rate and metallicity offsets}
\label{offset}

\begin{figure*}
	\includegraphics[width=\textwidth]{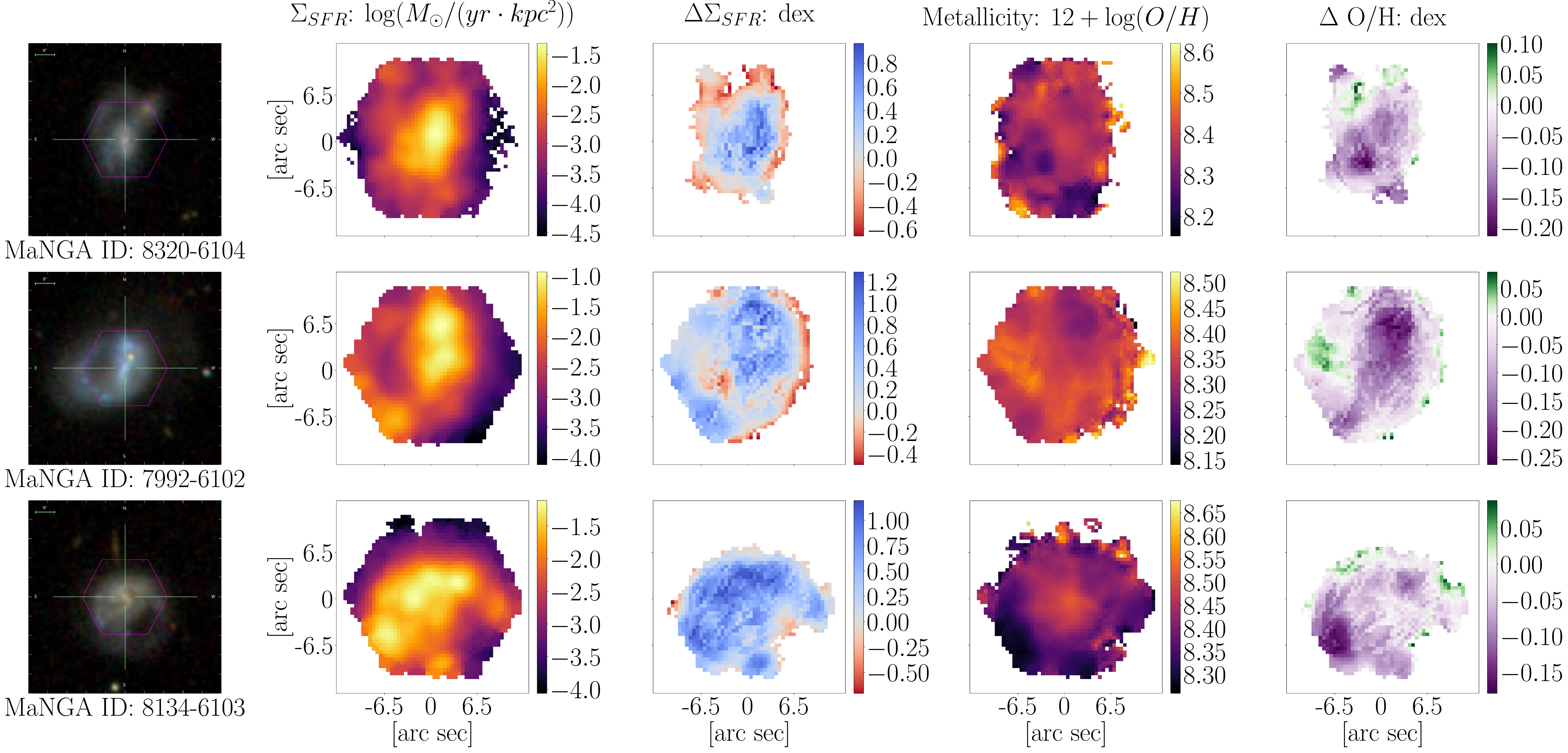}
    \caption{MaNGA data products and offset maps for 3 example post-merger galaxies. 1st column: SDSS $gri$ image with MaNGA IFU footprint overlaid in magenta. 2nd column: Map of \sigsfr\ as determined by {\sc pipe3d}. 3rd column: Offset in \sigsfr\ (\dsigsfr\ ) from the resolved main sequence; enhancements are shown in blue and deficits in red. 4th column: Map of O3N2 metallicity measurements from {\sc pipe3d}. 5th column: Offset in metallicity from  the resolved mass-metallicity relation ($\Delta$O/H); enhancements are shown in green and deficits in purple. Some spaxels are lost in the matching process for offset maps.} 
        \label{fig:maps}
\end{figure*}

To uncover the impact on SFR and O/H in post-merger galaxies due to the interaction, we need to establish a control sample against which to quantify enhancements and reductions in spaxel properties.  Due to the strong correlation between total stellar mass and both SFR and O/H (e.g. Brinchmann et al. 2004; Tremonti et al. 2004), it is already common procedure in merger studies to measure global offsets in SFR and O/H at fixed M$_{\star}$ (e.g. Ellison et al. 2008; Scudder et al. 2012; Patton et al. 2013).  The existence of `resolved' relations on the kpc-scale between \sigmass\ and both \sigsfr\  (S{\'a}nchez et al. 2013;  Cano-D{\'i}az et al. 2016; Gonz{\'a}lez-Delgado et al. 2016, Hsieh et al. 2017) and O/H (e.g. Rosales-Ortega et al. 2012; S{\'a}nchez et al. 2013; Barrera-Ballesteros et al. 2016), mean that a similar procedure can be followed for individual spaxels.  The spaxel offset method was first introduced, and described in detail, by Ellison et al. (2018) in their study of SFR profiles as a function of main sequence position, and we review only the main points here.

Following Ellison et al. (2018), we work only with spaxels that have a S/N of at least 3 in all relevant diagnostic emission lines, are classified as star-forming (Kauffmann et al. 2003b) and hence have reliable measurements of \sigmass, \sigsfr\, and O/H.  There are $\sim$20,000 and $\sim$900,000 spaxels in the 36 post-merger datacubes and full DR14, respectively, that satisfy these criteria.  For a given post-merger spaxel, control spaxels are then identified as those that match within 0.1 dex in \sigmass, 0.1 dex in $M_{\star}$, and 0.1 effective radii ($R_e$), where $r$-band $R_e$ is taken from Simard et al. (2011).  The median value of \sigsfr\ and O/H of the control sample is then subtracted from the post-merger spaxel value to yield \dsigsfr\ and $\Delta$O/H values on a spaxel-by-spaxel basis.  Three examples of our post-merger sample, with {\sc pipe3d} data products and maps of \dsigsfr\ and $\Delta$O/H, are presented in Fig. \ref{fig:maps}.  The median point spread function (PSF) of MaNGA observations is 2.5 arcsec, corresponding to 1.5 kpc at the typical redshift of the sample ($z=0.03$).

\section{Results \& Discussion}
\label{sec:results}

\subsection{Star formation rate profiles}

\begin{figure}
	\centering
	\begin{subfigure}{0.45\textwidth} 
	   \centering
		\includegraphics[width=\textwidth]{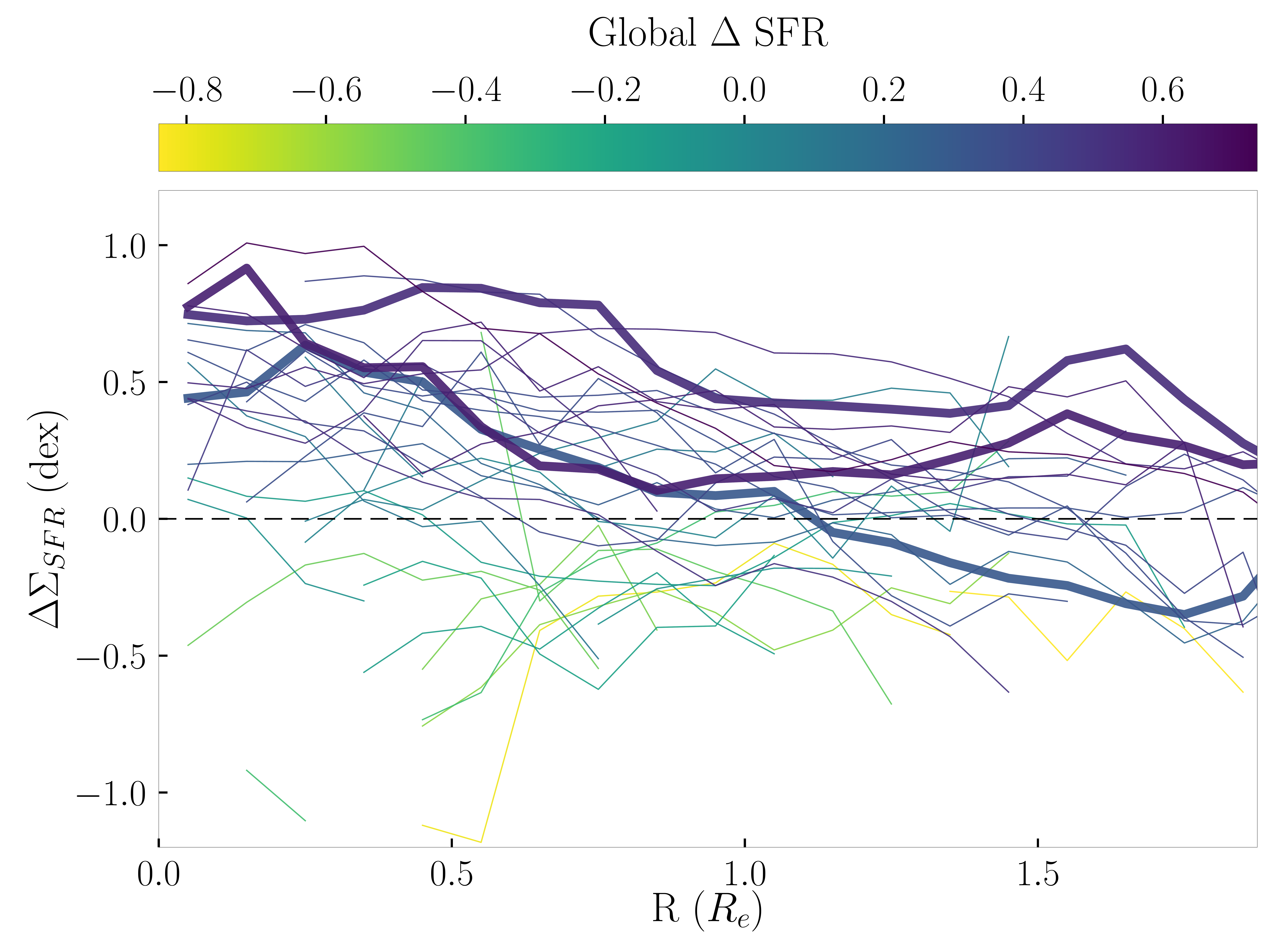}
	\end{subfigure}
	~
	\begin{subfigure}{0.45\textwidth} 
	    \centering
		\includegraphics[width=0.96\textwidth]{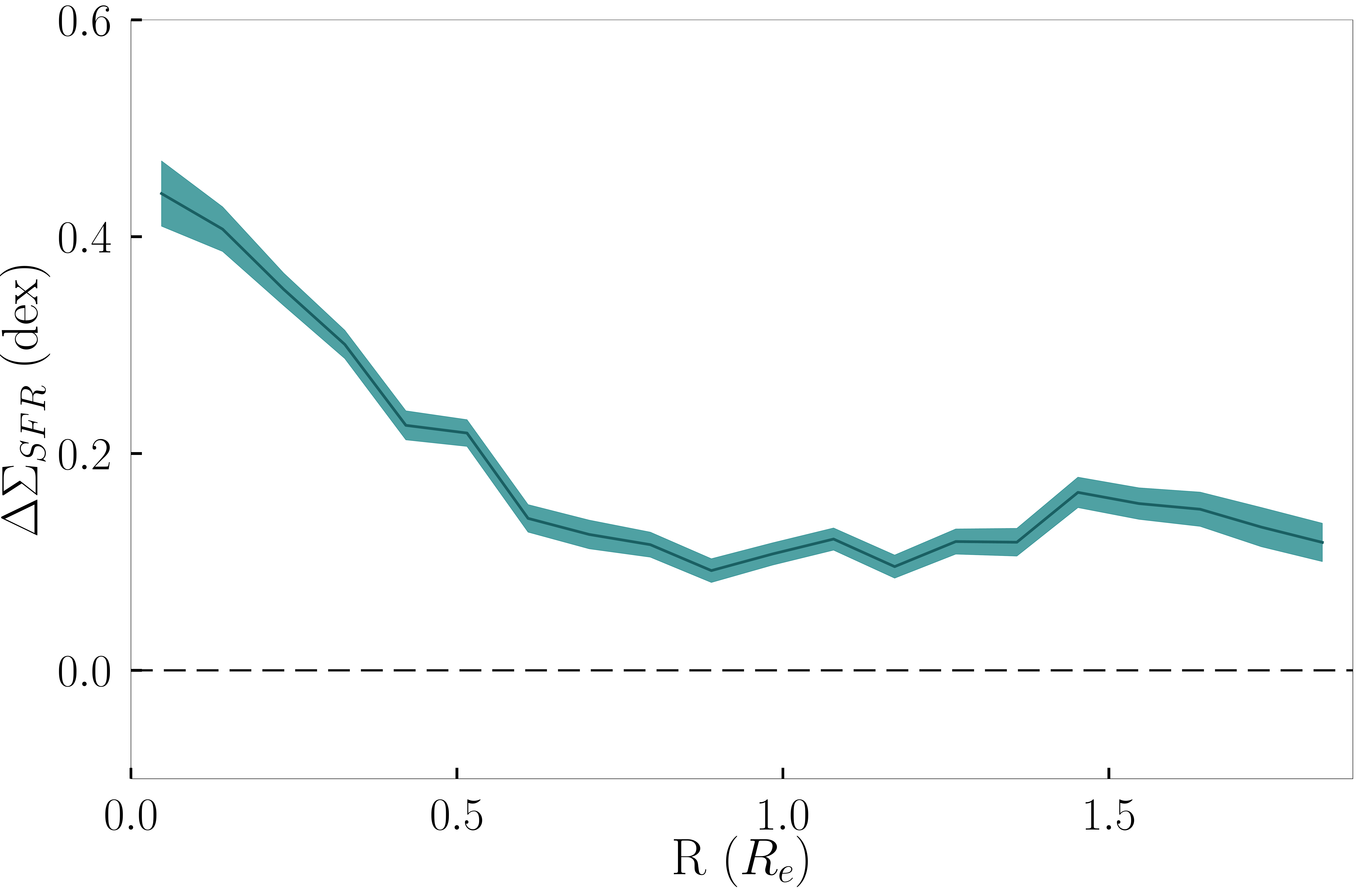}
	\end{subfigure}
	\caption{SFR enhancement in post-mergers as a function of radius. \textbf{Top panel:} All \dsigsfr\ profiles as a function of effective radius, coloured according to the global $\Delta$SFR of the host galaxy. Bold lines represent galaxies from Fig.\ref{fig:maps}.  \textbf{Bottom panel:} Median profiles for all galaxies, the width representing the error on the mean. Note that the top panel has a different y-scale.} 
	\label{fig:dSFR}
\end{figure}

In the top panel of Fig. \ref{fig:dSFR} we plot the radial profiles of \dsigsfr\ for each post-merger galaxy.  The individual gradients are colour-coded by their global $\Delta$SFR main sequence offsets. Incomplete gradients occur when there are incomplete \dsigsfr\ spaxel maps, which can result from, for example, contamination of the spaxel by an active galactic nucleus (AGN) or if there is insufficient S/N in the emission lines. The general trend found by Ellison et al. (2018), that galaxies with high global $\Delta$SFRs tend to have high \dsigsfr\ throughout, with a central peak in the SFR enhancement, is reproduced by the post-merger sample.  However, there is evidently a lot of variation amongst the post-merger gradients, particularly in the galactic outskirts. This is well illustrated by the three example post-mergers presented in Fig. \ref{fig:maps} whose \dsigsfr\ gradients are shown in bold in the top panel of Fig. \ref{fig:dSFR}.  Whilst all three galaxies from Fig. \ref{fig:maps} have elevated \dsigsfr\ in their inner regions, the post-merger example in the upper row has suppressed star formation in the outer disk, consistent with predictions by Moreno et al. (2015).  However, the example in the lower row of Fig. \ref{fig:maps} is enhanced throughout. The post-merger in the middle row of Fig. \ref{fig:maps} shows an asymmetric structure in its SFR enhancement, corresponding to extended streams and clumps in a tidal feature (middle left panel of Fig. \ref{fig:maps}). The top panel of Fig. \ref{fig:dSFR} also shows that 7 galaxies in the sample have suppressed \sigsfr\ in all radial bins where \sigsfr\ is measurable.

In the bottom panel of Fig. \ref{fig:dSFR} we combine all of the post-merger spaxels together to show the median \dsigsfr\ profile as a function of galactocentric radius in units of $R_e$. The profile is shown over a range in $R_e$ that includes 95 per cent of spaxels with measured \dsigsfr . The average post-merger profile shows a central \sigsfr\ enhancement of a factor of 2.5, with a modest 0.1 dex enhancement all the way to 1.9 $R_e$. This finding is consistent with previous hints from either smaller IFU studies (e.g. Garc{\'i}a-Mar{\'i}n et al. 2009; Cortijo-Ferrero et al. 2017) or from larger samples with limited spatial information (e.g. Ellison et al. 2013).  Indeed, the median \dsigsfr\ within a 1.5 arcsec radius of our post-merger sample is in excellent agreement with the single SDSS fibre $\Delta$SFR measured for post-mergers by Ellison et al. (2013).

However, our finding that the average post-merger profile has elevated star formation out to extended distances is apparently in contrast with the CALIFA sample of mergers studied by Barrera-Ballesteros et al. (2015) who found elevated central SFRs, but normal (or mildly suppressed) values over extended scales. There are several possible reasons for these different results.   The first difference is in sample selection. Where we have limited ourselves to a single evolutionary phase, namely recent (as evidenced by still-visible tidal features) post-mergers, in an attempt to restrict the timescale of the interactions under investigation, the CALIFA study samples mergers over a larger timescale, ranging from pairs with separations as large as $\sim$ 150 kpc to post-mergers.  The extraction of trends from this mixed sample could be complicated by different spatial distributions of star formation in the pre- and post-coalescence phases (e.g. Garc{\'i}a-Mar{\'i}n et al. 2009).  The second notable difference between these two works is in methodology.  Barrera-Ballesteros et al. (2015) bin their IFU data into two apertures in order to study central and extended star formation, whereas we construct full radial profiles. Finally, the CALIFA sample has a larger spatial coverage, out to 2.5 $R_e$.

In addition to finding that the average \dsigsfr\ profile is enhanced out to 1.9 $R_e$, we have shown that enhancements in star formation can be clumpy and offset from the centre of the galaxy (Fig. \ref{fig:maps}) and that sometimes the radial gradients can show suppressions of star formation, particularly in the outskirts (Fig. \ref{fig:dSFR}).  Again, both of these features have been hinted at from smaller studies (e.g. Elmegreen et al. 2006; Cortijo-Ferrero et al. 2017), although we have shown that spatially extended \sigsfr\ enhancements persist beyond the pair phase and are still measurable post-coalescence (c.f. Garc{\'i}a-Mar{\'i}n et al. 2009).  Simulations have also predicted both effects, from the combined influence of small scale compression and turbulence that triggers extended star formation (Powell et al. 2013; Renaud et al. 2014) and large scale gas inflows that may suppress star formation in the outer disk (Moreno et al. 2015).  The complexity revealed by our study motivates the future development of more sophisticated metrics, beyond azimuthally averaged radial profiles, to characterize spatial variations of \sigsfr\ in merging galaxies.

\subsection{Metallicity profiles}

\begin{figure}
	\centering
	\begin{subfigure}{0.45\textwidth} 
	    \centering
		\includegraphics[width=\textwidth]{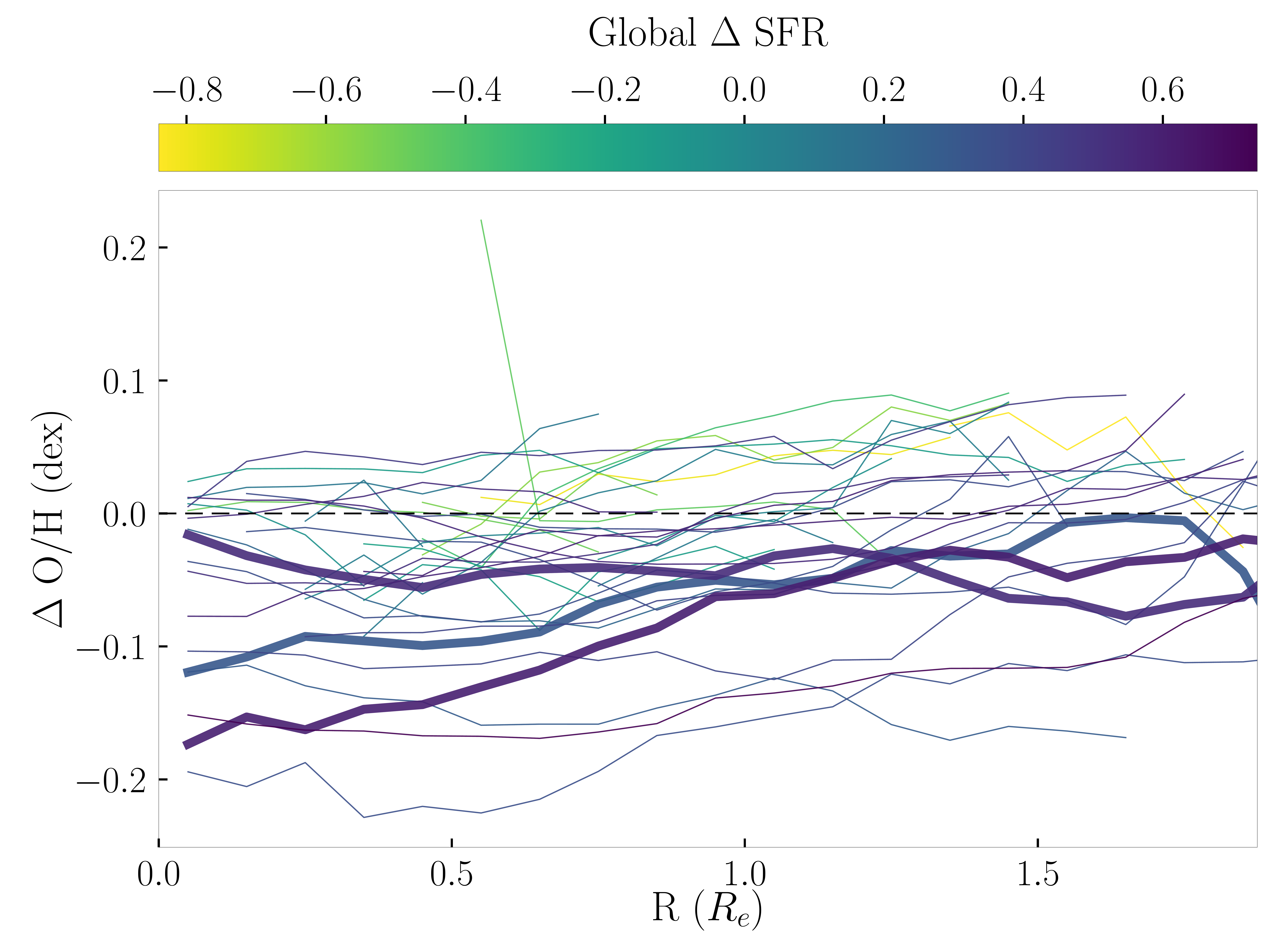}
	\end{subfigure}
	~
	\begin{subfigure}{0.45\textwidth} 
	    \centering
		\includegraphics[width=\textwidth]{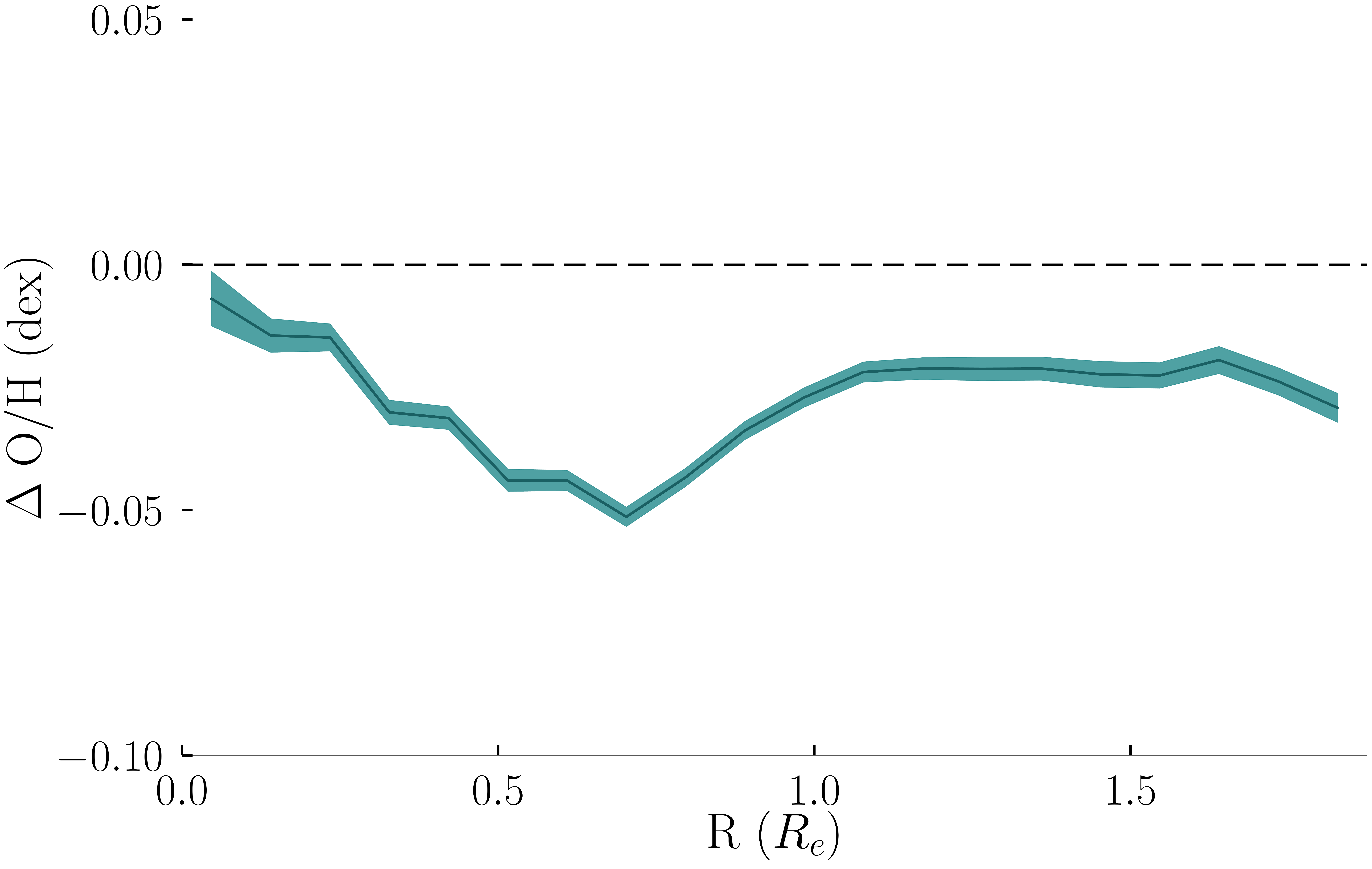}
	\end{subfigure}
		\caption{Metallicity suppression in post-mergers as a function of radius. \textbf{Top panel:} All $\Delta$O/H profiles plotted against effective radius, coloured according to the global $\Delta$SFR of the host galaxy. Bold lines represent galaxies from Fig.\ref{fig:maps}.  \textbf{Bottom panel:} Median profile for all galaxies, the width representing the error on the mean. Note that the top panel has a different y-scale.}
	\label{fig:do3n2}
\end{figure}

Fig.  \ref{fig:do3n2} presents results analogous to the \dsigsfr\  profiles in Fig. \ref{fig:dSFR}, but for gas phase metallicity. Although most of the post-mergers have either normal or suppressed central metallicities, there is again considerable variation from galaxy-to-galaxy in the outskirts. The spaxel median yields values that are suppressed on extended scales throughout the disk, to at least 1.9 $R_e$, by $\sim -0.04$ dex on average. Although the uncertainties of individual spaxel metallicities are $\sim -0.1$ dex (Marino et al. 2013), the statistical error on the median spaxel profile is very small, and the median offset therefore significant, thanks to the very large number of spaxels ($\sim$1000) in each radial bin.

A number of observational (e.g. Ellison et al. 2008; Michel-Dansac et al. 2008; Scudder et al. 2012) and simulation-based (e.g. Montuori et al. 2010; Torrey et al. 2012; Bustamante et al. 2018) studies have established that central metallicities are diluted as the result of galaxy interactions.  However, relatively few works have studied the spatial extent of this dilution.  Simulations that have investigated the changing O/H gradient during mergers have found that dilution is quite widespread, occurring out to 2--3 disk scale lengths (Rupke et al. 2010a; Perez et al. 2011; Sillero et al. 2017), or $\sim$ 6--8 kpc, in good agreement with the extent of the metallicity offsets seen in Fig. \ref{fig:do3n2}.  The simulations also predict that metallicities are enhanced on larger scales (the combination of which, with inner dilutions, leads to the overall flatter gradients).  Unfortunately, our data do not extend to sufficiently large radii to test this prediction with robust statistics. Intriguingly, Fig. \ref{fig:do3n2} indicates the smallest level of metallicity dilution occurs at the smallest radii, where the SFR offsets are the greatest (Fig. \ref{fig:dSFR}).  Although the small $\Delta$O/H at small radii could potentially be the result of re-enrichment of gas at small radii, as a result of a starburst, we note that this effect is seen at approximately the spatial resolution limit of the data.

As for our SFR results, the widespread metallicity dilution seen in the MaNGA post-merger sample is complementary to observations of flattened abundance gradients in mergers (e.g. Kewley et al. 2010; Rupke et al. 2010b; Rich et al. 2012, S{\'a}nchez et al. 2014), which both support large-scale metal-poor gas inflows and agree with the galaxy-wide O/H dilution seen in MaNGA maps of asymmetric galaxies (Rowlands et al. 2018).  However, again our results are in contrast to the CALIFA merger study of Barrera-Ballesteros et al. (2015) -- whilst the metallicities are suppressed in the outskirts of CALIFA mergers, they are normal in the central regions. Again, discrepancies between these two works likely stem from a combination of different sample selection and methodologies, particularly the inclusion of interacting galaxies at any stage.

\section{Conclusions}

This is the first study to derive full radial profile information in any statistically significant post-merger IFU sample. We have quantified enhancements/deficits in SFR and metallicity for $\sim$20,000 spaxels in 36 visually classified post-merger galaxies in MaNGA, by computing offsets from the relations of O/H and \sigsfr\ versus \sigmass.  On average, we find that post-mergers have central \sigsfr\ enhancements of a factor $\sim$ 2.5, but with low level enhancements, by $\sim$ 0.1 dex, out to at least 1.9 $R_e$  (Fig. \ref{fig:dSFR}). However, whilst elevated central SFRs are the `norm' amongst post-mergers, we find that both suppressions and enhancements can occur in the outer regions. On average the metallicities are diluted out to at least 1.9 $R_e$, typically suppressed by $\sim -0.04$ dex (Fig. \ref{fig:do3n2}). The full 2D maps of \dsigsfr\ and $\Delta$O/H (Fig. \ref{fig:maps}) reveal that star formation can be clumpy, offset from the centre, and in non-axisymmetric tidal features.   Our results indicate that radial profiles are insufficient to capture the full diversity of responses to the interaction and motivate the development of a more complex 2D metric to map \dsigsfr\ and $\Delta$O/H, which we will present in a future work.

\section*{Acknowledgements}

The authors thank Trystyn Berg, Maan Hani, Connor Bottrell, Joanna Woo, Jillian Scudder, Jorge Barrera-Ballesteros, Lihwai Lin, and the anonymous referee for their insightful comments on the works herein. This project makes use of the MaNGA-Pipe3D dataproducts. We thank the IA-UNAM MaNGA team for creating this catalogue, and the ConaCyt-180125 project for supporting them. SFS is grateful for the support of a CONACYT (Mexico) grant CB-285080, and funding from the PAPIIT-DGAPA-IA101217(UNAM).

Funding for the Sloan Digital Sky Survey IV has been provided by the Alfred P. Sloan Foundation, the U.S. Department of Energy Office of Science, and the Participating Institutions. SDSS-IV acknowledges
support and resources from the Center for High-Performance Computing at
the University of Utah. The SDSS web site is www.sdss.org.

SDSS-IV is managed by the Astrophysical Research Consortium for the 
Participating Institutions of the SDSS Collaboration including the 
Brazilian Participation Group, the Carnegie Institution for Science, 
Carnegie Mellon University, the Chilean Participation Group, the French Participation Group, Harvard-Smithsonian Center for Astrophysics, 
Instituto de Astrof\'isica de Canarias, The Johns Hopkins University, 
Kavli Institute for the Physics and Mathematics of the Universe (IPMU) / 
University of Tokyo, Lawrence Berkeley National Laboratory, 
Leibniz Institut f\"ur Astrophysik Potsdam (AIP),  
Max-Planck-Institut f\"ur Astronomie (MPIA Heidelberg), 
Max-Planck-Institut f\"ur Astrophysik (MPA Garching), 
Max-Planck-Institut f\"ur Extraterrestrische Physik (MPE), 
National Astronomical Observatories of China, New Mexico State University, 
New York University, University of Notre Dame, 
Observat\'ario Nacional / MCTI, The Ohio State University, 
Pennsylvania State University, Shanghai Astronomical Observatory, 
United Kingdom Participation Group,
Universidad Nacional Aut\'onoma de M\'exico, University of Arizona, 
University of Colorado Boulder, University of Oxford, University of Portsmouth, 
University of Utah, University of Virginia, University of Washington, University of Wisconsin, 
Vanderbilt University, and Yale University.









\bsp	
\label{lastpage}
\end{document}